\documentclass[11pt]{article}

\usepackage{amsthm, amsfonts}
\usepackage[margin=2.5cm]{geometry}
\usepackage{amssymb}
\usepackage[]{amsmath}
\usepackage{fancyhdr}
\usepackage[cp1250]{inputenc}

\newtheorem{thm}{Theorem}[section]

\usepackage{hyperref}
\hypersetup{
    colorlinks=true,
    linkcolor=blue,
    filecolor=magenta,      
    urlcolor=cyan,
}

\title{Dynamic Quantum Tomography Model for Phase-Damping Channels}
\author{\Large Artur Czerwi\'nski, Andrzej Jamio\l{}kowski\\
Institute of Physics, Faculty of Physics, Astronomy and Informatics,\\
Nicolaus Copernicus University,\\
Grudziadzka 5, 87-100 Torun, Poland\\
Corresponding author's e-mail:  aczerwin@fizyka.umk.pl
}
\begin{document}
\maketitle

\abstract{In this article we propose a dynamic quantum tomography model for open quantum systems with evolution given by phase-damping channels. Mathematically, these channels correspond to completely positive trace-preserving maps defined by the Hadamard product of the initial density matrix with a time-dependent matrix which carries the knowledge about the evolution. Physically, there is a strong motivation for considering this kind of evolution because such channels appear naturally in the theory of open quantum systems. The main idea behind a dynamic approach to quantum tomography claims that by performing the same kind of measurement at some time instants one can obtain new data for state reconstruction. Thus, this approach leads to a decrease in the number of distinct observables which are required for quantum tomography; however, the exact benefit for employing the dynamic approach depends strictly on how the quantum system evolves in time. Algebraic analysis of phase-damping channels allows one to determine criteria for quantum tomography of systems in question. General theorems and observations presented in the paper are accompanied by a specific example, which shows step by step how the theory works. The results introduced in this article can potentially be applied in experiments where there is a tendency a look at quantum tomography from the point of view of economy of measurements, because each distinct kind of measurement requires, in general, preparing a separate setup.
}

\section{Introduction}\label{sec1}

Quantum tomography initiated in 1933 when Pauli posted a problem whether the quantum wavefunction of a physical system which is assumed to be pure can be uniquely determined by its position and momentum probability distributions \cite{pauli80,reich44}. Currently it is commonly known that in general Pauli's problem is not uniquely solvable for any wavefunction \cite{reich44,corbett06}. Gerchberg-Saxton algorithm is one of the tools that allow one to compute the quantum wavefuntion when it is feasible \cite{gerchberg72}. This algorithm has been widely applied in all areas of science where the problem of phase retrieval occurs. Since 1933 there have been proposed many other approaches to quantum wavefunction reconstruction. For instance, in 2011 a group of researchers demonstrated that the quantum wavefunction can be measured in a direct way by means of the weak measurement\cite{bamber11}. Since 2011 their approach has received much attention and many other tomography models based on weak measurement have been proposed. Some consider weak measurement as a tool to increase efficacy of quantum tomography, whereas others look at this approach more critically \cite{gross15}.

The term quantum tomography refers not only to the problem of wavefunction reconstruction, but to any method which aims to reconstruct the accurate representation of a quantum system on the basis of data obtainable from an experiment. Thus, the term quantum tomography is used also in reference to Wigner function or density matrix reconstruction -- this article focuses on the latter. According to the assumptions of quantum mechanics the density matrix contains all achievable information about a physical system. If $\mathcal{H}$ denotes a finite dimensional Hilbert space ($dim\mathcal{H} = n$) associated with the physical system, by $\mathcal{S(H)}$ we shall denote the state set, i.e. the set of all legitimate density matrices: $\mathcal{S(H)} = \{\rho: \mathcal{H} \rightarrow \mathcal{H}, \rho \geq 0, Tr \rho =1\}$. Additionally, we shall use $\mathcal{B(H)}$ in reference to the space of all linear operators on $\mathcal{H}$ and $\mathcal{B_* (H)}$ for the space of all Hermitian operators on $\mathcal{H}$.

Some fundamental results concerning quantum state tomography can be found in \cite{altepeter04}. In another paper the authors proposed a method of reconstructing the density matrix from projections \cite{ole95}. Also, the approach to quantum tomography based on the weak measurement has been generalized so that it can be applied to density matrix reconstruction as well \cite{bamber12,wu13}.

In this article we refer to yet another approach to density matrix reconstruction which was introduced in 1983 in \cite{jam83}. It is the so-called stroboscopic tomography which employs the knowledge about the evolution of a quantum system in order to determine the optimal criteria for quantum tomography. This approach was developed in many subsequent papers such as \cite{jam00,jam04,jam12}. In the original formulation of the stroboscopic tomography the underlying assumption claims that the evolution of a quantum system is given by a master equation of the form
\begin{equation}\label{1}
\frac{d \rho}{d t} = \mathbb{L}[\rho],
\end{equation}
where $\mathbb{L}$ is a linear operator (time-independent) and is referred to as the generator of evolution.

Naturally during the evolution the density matrix cannot leave the state set, i.e. $\forall_{t>0} \text{ } \rho(t) \in \mathcal{S(H)}$. The most general form of a legitimate generator of evolution, which preserves trace and positivity, can be written in the diagonalized form as  \cite{gorini76,lindblad76}:
\begin{equation}\label{1.1}
\mathbb{L} [\rho] = - i [H,\rho] + \frac{1}{2} \sum_{i=1} ^{N^2 -1} \gamma_ i \left ( [V_i \rho, V_i ^*] + [V_i, \rho V_i ^*]  \right ),
\end{equation}
where $H\in  \mathcal{B_*(H)}$ and $\gamma_i \geq 0 $. Operators $V_i \in \mathcal{B(H)}$ are called Lindblad operators.

The evolution equation \eqref{1} gives one the formula for $\rho(t)$ for any $t \in \mathbb{R}_+$ by a semigroup
\begin{equation}\label{2}
\rho(t) = exp(\mathbb{L}t)\rho(0),
\end{equation}
where $exp(\mathbb{L}t)$ can be expanded by means of a finite number of operators if one introduces the notion of the minimal polynomial. The ability to determine the polynomial representation of the semigroup plays an important role in the stroboscopic tomography \cite{jam04}.

In the stroboscopic tomography there are two main assumptions.\\
1. one can measure physical quantities associated with Hermitian operators from a fixed set $\{Q_1, \dots, Q_r\}$ $(r< n^2-1)$, which is not informationally complete\\
2. one possesses the knowledge about evolution of the system encoded in the equation \eqref{1}.

The set of observables $\{Q_1, \dots, Q_r\}$ is not complete which means that it does not satisfy the four equivalent definitions of spanning set presented in \cite{ariano2000}. Therefore, single measurement of each observable does not provide us with knowledge sufficient to determine the intial density matrix.

However, knowledge about evolution makes it possible to observe the system on an interval $[0,T]$. Then, if the number of distinct observables is not lower than the index of cyclicity of the generator $\mathbb{L}$,  there exists a sequence of time instants $0\leq t_1 < t_2 < \dots < t_p \leq T$ such that the initial density matrix can be computed from the data $m_i (t_j) = Tr(Q_i \rho(t_j))$ where $i=1,\dots, r$ and $j=1,\dots, p$. Obviously, we assume that each measurement is performed on a distinct copy of the system. Thus, to employ the stroboscopic tomography one has to be able to prepare a certain number of systems in identical quantum state. To observe how the stroboscopic tomography works for a specific example one can refer to \cite{czerwin15}. In order to learn how in general for some evolution models this approach can improve the process of state reconstruction one can refer to a review paper \cite{jam12} which contains all fundamental results concerning optimal criteria for quantum tomography. In case of phase-damping channels, which are discussed in this paper, we explain in details in section \ref{sec3} why it is beneficial to employ the idea of repeated measurement of a fixed set of observables in the context of quantum tomography.

In this article we employ the general idea of the stroboscopic tomography to phase-damping channels (pure decoherence) \cite{havel01,helm11}. The main difference between the known results of the stroboscopic tomography and the current paper is that here we analyze a subclass of completely positive and trace-preserving maps given by the Hadamard product of the initial density matrix and a time-dependent positive definite matrix. The dynamical map from equation \eqref{2} in the current analysis is substituted by the form
\begin{equation}\label{3}
\rho(t) = D(t) \circ \rho(0),
\end{equation}
where $D(t) \in \mathbb{M}_n(\mathbb{C})$ and $n = dim \mathcal{H}$. The necessary and sufficient conditions for the equation \eqref{3} to describe evolution of a quantum system are:\\
$1. \hspace{5pt} \forall_{t>0} \hspace{10pt} D(t) \geq 0$ (condition for complete positivity),\\
$2. \hspace{5pt} d_{ii} (t) =1$  for  $i=1,\dots, n$  (condition for trace-preservity),\\
$3. \hspace{5pt}  d_{ij}(0) =1$ for $i,j=1,\dots, n$  (initial condition).

One can observe that the above conditions ensure that the map given by \eqref{3} describes a legitimate evolution of a physical system. One can also notice that taking as the starting point the map from \eqref{3} we can consider more general cases than with the evolution equation \eqref{1}, because depending on the structure of $D(t)$ the map from \eqref{3} may correspond to evolution equations with generator either time-dependent or time-independent.

In case of stroboscopic tomography as introduced in \cite{jam83} one needs to use algebraic properties of a quantum semigroup given by \eqref{2}. Whereas in the current article we need to employ different algebraic methods to consider the dynamical map as introduced in \eqref{3}. Therefore, though seemingly the current article resembles previous works on stroboscopic tomography \cite{jam83,jam00,jam04,jam12,czerwin15}, it actually brings a significant contribution to the field because it differs with the main assumption concerning evolution and mathematical methods used to solve this problem. Morover, the current approach can be applied to more general evolution models with time-dependent generators. In some specific cases it might be better to follow the approach as introduced in \cite{jam83}, whereas in other cases it might be more beneficial to employ the current reasoning. More about the usefulness of the current approach and possible problems that might arise when using it can be found in sections \ref{sec3} and \ref{sec4}.

This article, apart from the introduction, consists of three main parts. In section \ref{sec2} we give more physical motivation for research into the quantum channels of the form \eqref{3}. Then in section \ref{sec3} we introduce general concepts about quantum tomography for phase-damping channels. Finally in sections \ref{sec4} and \ref{sec5} we demonstrate two specific examples of quantum tomography for phase-damping channels.

\section{Phase-Damping Channels}\label{sec2}

The dynamics of an open quantum system, i.e. a physical system $\mathcal{S}$ described by $\rho(t) \in \mathcal{S}(\mathcal{H}_S)$ interacting with an environment $\mathcal{E}$ described by quantum state $\rho_E (t) \in \mathcal{S}(\mathcal{H}_E)$, has been the subject of many books \cite{breuer07,alicki07} and articles \cite{chruscinski14}. Obviously, the evolution of the total system $\mathcal{S}+\mathcal{E}$ (described by the density matrix $\rho_{SE} (t)$) is unitary and determined by a Hamiltonian given as
\begin{equation}\label{4}
H = H_S \otimes \mathbb{I}_E + \mathbb{I}_S \otimes H_E + H_{int}, 
\end{equation}
where $H_S, \mathbb{I}_S : \mathcal{H}_S \rightarrow \mathcal{H}_S$, $H_E, \mathbb{I}_E : \mathcal{H}_E \rightarrow \mathcal{H}_E$ and $H_{int}:  \mathcal{H}_S \otimes  \mathcal{H}_E \rightarrow \mathcal{H}_S \otimes  \mathcal{H}_E$.

Under two approximations one is able to derive a computable formula for evolution of the quantum system of interest. One crucial assumptions claims that the coupling between the system and the environment is weak and, therefore, the quantum state of the environment does not change in time and can be denoted simply by $\rho_E$. The other important assumption claims that there is no initial correlation between the system and its environment, i.e. $\rho_{SE} (0) = \rho(0) \otimes \rho_{E}$. Under such two assumptions the dynamical map given by means of the partial trace over the environmental degrees of freedom has the form
\begin{equation}\label{5}
\rho (t) = Tr_{E} \left( U(t) \rho(0) \otimes \rho_E U^* (t) \right ) \equiv \Lambda_t [\rho(0)],
\end{equation}
where $U(t)$ is the unitary operator that governs the evolution of the total system $\mathcal{S}+\mathcal{E}$, i.e. $U(t) = exp(-i H t)$.

To introduce the model of pure decoherence we shall add more assumptions concerning the Hamiltonian from \eqref{4}. First, let us assume that the eigenvectors and eigenvalues of $H_S$ are known, i.e. one has the equations
\begin{equation}\label{6}
H_S |n\rangle = e_n |n\rangle \text{ for } n=1,2, \dots, dim\mathcal{H}_S.
\end{equation}

The eigenbasis of $H_S$ is then used to define the interaction Hamiltonian $H_{int}$:
\begin{equation}\label{7}
H_{int} = \sum_{n} \mathcal{P}_n \otimes B_n,
\end{equation}

where $ \mathcal{P}_n = |n \rangle \langle n|$ and $B_n: \mathcal{H}_E  \rightarrow \mathcal{H}_E$. Then the full Hamiltonian can be rewritten as
\begin{equation}\label{8}
\begin{aligned}
H  {} & = \sum_{n} e_n \mathcal{P}_n \otimes \mathbb{I}_E + \sum_{n} \mathcal{P}_n \otimes H_E +  \sum_{n} \mathcal{P}_n \otimes B_n = \\&
= \sum_{n} \mathcal{P}_n \otimes \left( e_n \mathbb{I}_E + H_E + B_n \right ) = \\&
= \sum_{n} \mathcal{P}_n \otimes Z_n,
\end{aligned}
\end{equation}
where $Z_n \equiv e_n \mathbb{I}_E + H_E + B_n $. One can observe that this Hamiltonian possesses the following property
\begin{equation}\label{9}
H^k = \sum_{n} \mathcal{P}_n \otimes Z_n ^k,
\end{equation}
where $k=1,2,\dots$ By employing this property one can easily obtain a formula for the unitary operator corresponding to the full Hamiltonian
\begin{equation}\label{10}
U(t) = exp(-i H t) = \sum_{n} \mathcal{P}_n \otimes e^{- i Z_n t}.
\end{equation}

Now we shall substitute the formula \eqref{10} to the general equation for a dynamical map \eqref{5} and obtain
\begin{equation}\label{11}
\begin{aligned}
\Lambda_t [\rho(0)] {}& = Tr_E \left\{ \sum_{n} \mathcal{P}_n \otimes e^{- i Z_n t} \rho(0) \otimes \rho_E \sum_{m} \mathcal{P}_n \otimes e^{ i Z_m t} \right\} = \\&
= \sum_{n,m} Tr_E \left\{ \mathcal{P}_n \rho(0) \mathcal{P}_m  \otimes e^{- i Z_n t} \rho_E e^{ i Z_m t} \right\} = \\&
= \sum_{n,m} \mathcal{P}_n \rho(0) \mathcal{P}_m Tr \left( e^{- i Z_n t} \rho_E e^{ i Z_m t} \right) = \\&
= \sum_{n,m} C_{nm} (t) \mathcal{P}_n \rho(0) \mathcal{P}_m,
\end{aligned}
\end{equation}
where $C_{nm} (t) \equiv Tr \left( e^{- i Z_n t} \rho_E e^{ i Z_m t} \right)$. Apparently, the final formula for the dynamical map is simply the Kraus form of a completely positive map, from which we can observe that $\forall_{t>0}$ $[C_{nm} (t)] \geq 0$. The dynamical map can also be represented in a different way -- if one decomposes the initial density matrix $\rho(0)$ in the eigenbasis of $H_S$ introduced in \eqref{6}, i.e.
\begin{equation}\label{12}
\rho(0) = \sum_{i,j} \rho_{ij} |i \rangle \langle j| \text{ for } i,j=1, \dots, dim\mathcal{H}_S,
\end{equation}
one can get
\begin{equation}\label{13}
\begin{aligned}
\rho_{ij} (t) {}& = \langle i | \rho(t)| j\rangle  = \langle i | \Lambda_t [\rho(0)] | j\rangle= \langle i | \sum_{n,m} C_{nm} (t) \mathcal{P}_n \rho(0) \mathcal{P}_m |j\rangle =\\& = \sum_{n,m} C_{nm} (t) \langle i | \mathcal{P}_n \rho(0) \mathcal{P}_m | j \rangle = C_{ij} (t) \langle i | \rho(0)| j\rangle= C_{ij} (t) \rho_{ij},
\end{aligned}
\end{equation}
which allows to rewrite the formula for the dynamical map by means of the Hadamard product
\begin{equation}\label{14}
\rho (t) = [C_{ij} (t) ] \circ \rho(0).
\end{equation}

One can also calculate that
\begin{equation}\label{15}
C_{jj} = Tr \left( e^{-i Z_j t} \rho_E e^{i Z_j t} \right) = Tr\rho_E = 1
\end{equation}
for $j=1,\dots, dim\mathcal{H}_S$. Obviously, the equation \eqref{15} ensures that the map \eqref{14} is trace-preserving.

Furthermore, one can instantly notice that 
\begin{equation}\label{16}
C_{ij} (0) = Tr \rho_E = 1,
\end{equation}
which ensures that the initial condition $\Lambda_0 [\rho(0)] = \rho(0)$ is fulfilled.

From the analysis presented in this section one can conclude that there is explicit physical motivation for considering dynamical maps as introduced in \eqref{3}. Such dynamical maps expressed by the Hadamard product appear naturally in the theory of open quantum systems if one makes the assumptions about the Hamiltonian as \eqref{6}-\eqref{7}. This kind of evolution model of open quantum systems is commonly referred to as pure decoherence or phase-damping channels. Therefore, it seems utterly justifiable to formulate questions concerning criteria for optimal tomography of systems with evolution given by \eqref{3}.

\section{Quantum Tomography for Phase-Damping Channels}\label{sec3}

In this section we will be analyzing criteria for quantum tomography of an open quantum system with evolution given by 
\begin{equation}\label{17}
\rho(t) = D(t) \circ \rho(0),
\end{equation}
where $D(t)$ satisfies the conditions enumerated in section \ref{sec1}. The goal of quantum tomography is to determine the initial density matrix $\rho(0)$ on the basis of data obtainable from an experiment. We assume to have a set of observables $\{Q_1, \dots, Q_r\}$ and each of them can be measured at discrete time instants $\{t_1, \dots, t_p\}$. From physical point of view it appears more natural to assume that from an experiment we can obtain a discrete rather than continuous data. Thus, from an experiment we obtain a matrix of data $[m_i(t_j)]$ and the elements of this matrix can be expressed as
\begin{equation}\label{18}
m_i (t_j) = Tr\left\{ Q_i (D(t_j) \circ \rho(0))\right\}.
\end{equation}

The crucial idea that will be used to trasform the equation for a measurement result claims that for any continuous time-dependent matrix $D(t) \in \mathbb{M}_n (\mathbb{C})$ there exist a set of $\mu$ linearly independent matrices $A_k \in \mathbb{M}_n (\mathbb{C})$ and a set of $\mu$ time-dependent and linearly independent functions $\lambda_k (t): \mathbb{R} \rightarrow \mathbb{C}$ such that the matrix $D(t)$ can be decomposed as
\begin{equation}\label{decom}
D(t) = \sum_{k=1}^{\mu} \lambda_k (t) A_k.
\end{equation}
In other words we shall say that the set $\{A_1, \dots, A_{\mu}\}$ constitutes a constant basis for the time-dependent matrix $D(t)$. To observe how such decomposition can be obtained, one can take $t_1 \geq 0$ such that $D(t_1) \neq 0$ and set $A_1 \equiv D(t_1)$. Then, if there exists a function $\lambda_1(t)$ such that $D(t) = \lambda_1 (t ) A_1$, then the decomposition has been achieved. If no such function $\lambda_1 (t)$ can be found, there exists a $t_2 \geq 0$ such that $A_2  := D(t_2) \neq \lambda_1 (t) A_1$ for any scalar $\lambda_1 (t)$. Now, either there exist scalar functions $\lambda_1(t)$ and $\lambda_2 (t)$ such that we have $D(t)  = \lambda_1(t) A_1 + \lambda_2(t) A_2$ or there exists a $t_3$ such that with $A_3 := D(t_3)$ the set $\{ A_1, A_2, A_3\}$ is linearly independent. Clearly, this procedure can be continued as necessary and has to terminate after $\mu$ steps, where $\mu \leq n^2$. Finally, we obtain a set of $\mu$ constant matrices $A_1, \dots, A_{\mu}$ such that the equality \eqref{decom} holds for some scalar functions $\lambda_i (t)$, $i=1,\dots, \mu$.

There is another possible situation -- at the beginning one may know a decomposition of the matrix $D(t)$ in a form:
\begin{equation}
D(t) = \sum_{j=1}^m \beta_j (t) B_j,
\end{equation}
where $B_j$, $j=1, \dots,m$, are some constant matrices. Initially, neither the matrices $B_j$ nor the functions $\beta_j (t)$ have to be linearly independent. Then one take a subset $\{B_1, \dots, B_k\}$ containing the matrices which are linearly independent and next one calculates all possible products $B_i B_h$, where $i,h = 1, \dots, k$. Then one supplements the set $\{B_1, \dots, B_k\}$ by new linearly independent results of these multiplications. After a finite number of repetitions the initial base $B_j$ where $j=1, \dots,m$ is substituted by linearly independent set $B'_1, \dots, B'_{\mu}$, which can be used to present $D(t)$ in the desired form \eqref{decom}.

Substituting $D(t) = \sum_{k=1}^{\mu} \lambda_k (t) A_k$ to \eqref{18} one gets
\begin{equation}\label{19}
m_i (t_j) = \sum_{k=1}^{\mu} \lambda_k (t_j) Tr\left\{ Q_i (A_k \circ \rho(0))\right\}.
\end{equation}

Now, we shall recall a theorem that indicates a connection between the Hadamard product and the standard matrix product \cite{schott05}.
\begin{thm}\label{thm:1}
Let $A,B$ and $C$ be any $n \times m$ dimensional matrices. Then the following equality holds
\begin{equation}\label{20}
Tr\{ A^T (B  \circ C)\} = Tr\{(A^T \circ B^T) C\}
\end{equation}
\end{thm}

To observe that the equality \eqref{20} holds one can take $A = [a_{ij}], B=[b_{ij}]$ and $C=[c_{ij}]$ and then perform simple calculations to compare the left-hand side and right-hand side of the equation \eqref{20}.

Here, in case of equation \eqref{19} the theorem \ref{thm:1} allows one to rewrite the formula for a measurement result in a convenient way
\begin{equation}\label{21}
m_i (t_j) = \sum_{k=1}^{\mu} \lambda_k (t_j) Tr\left\{ (Q_i \circ A_k^T) \rho(0) \right\}.
\end{equation}

One can notice that if the measurement of any observable $Q_i$ is performed at distinct time instants $t_1,\dots, t_p$ we obtain a set of $p$ equations:

\begin{equation}\label{22}
\begin{aligned}
{} & m_i (t_1) = \sum_{k=1}^{\mu} \lambda_k (t_1) Tr\left\{ (Q_i \circ A_k^T) \rho(0) \right\}, \\ &
m_i (t_2) = \sum_{k=1}^{\mu} \lambda_k (t_2) Tr\left\{ (Q_i \circ A_k^T) \rho(0) \right\}, \\ &
\vdots \\&
m_i (t_p) = \sum_{k=1}^{\mu} \lambda_k (t_p) Tr\left\{ (Q_i \circ A_k^T) \rho(0) \right\}. 
\end{aligned}
\end{equation}

One can notice that such a system of equations can be rewritten as a matrix equation
\begin{equation}\label{23}
\left[ \begin{matrix} m_i (t_1) \\ m_i (t_2) \\ \vdots \\ m_i (t_p) \end{matrix} \right] = 
\left[\begin{matrix}\lambda_1 (t_1) & \lambda_2 (t_1) & \dots & \lambda_{\mu} (t_1)     \\ \lambda_1 (t_2) & \lambda_2 (t_2) & \dots & \lambda_{\mu} (t_2) \\ 
\vdots & \vdots & \ddots & \vdots \\ 
\lambda_1 (t_p) &  \lambda_2 (t_p) & \dots & \lambda_{\mu} (t_p)
\end{matrix} \right] \left[ \begin{matrix}Tr\left\{ (Q_i \circ A_1 ^T) \rho(0) \right\}\\ Tr\left\{ (Q_i \circ A_2^T) \rho(0) \right\} \\ \vdots \\  Tr\left\{ (Q_i \circ A_{\mu}^T) \rho(0) \right\} \end{matrix} \right].
\end{equation}

On the left-hand side of the matrix equation \eqref{23} one has the vector of data which is accessible from an experiment, whereas on the right-hand side one has the matrix $[\lambda_k (t_j)]$  which is computable on the basis of the structure of $D(t)$. Thus, one can formulate a theorem concerning the condition for computability of the projections $Tr\left\{ (Q_i \circ A_k^T) \rho(0) \right\}$ ($k=1,\dots,\mu$) from the equation \eqref{23}.

\begin{thm}\label{thm2}
The equation \eqref{23} allows one to compute the projections $Tr\left\{ (Q_i \circ A_k^T) \rho(0) \right\}$ where $k=1,\dots,\mu$ if and only if

\begin{equation}\label{24a}
rank [\lambda_k (t_j)] = \mu
\end{equation}

Evidently, the condition \eqref{24a} claims that the matrix $[\lambda_k (t_j)]$ has to have full rank, which means that the number of measurements cannot be lower than $\mu$, i.e. $p \geq \mu$.
\end{thm}

Naturally, from physical point of view it is desired to perform quantum tomography with the lowest possible number of distinct measurements. Therefore, the optimal form of the equation \ref{23} is when the matrix $[\lambda_k (t_j)]$ is square. Then the projections $Tr\left\{ (Q_i \circ A_k^T) \rho(0) \right\}$ ($k=1,\dots,\mu$) are computable iff  $det[\lambda_k (t_j)] \neq 0$.

One can observe that by performing the measurement of the same observable $Q_i$ at distinct time instants $\{t_1,\dots, t_p\}$ we are able to find $Tr\left\{ (Q_i \circ A_k^T) \rho(0) \right\}$ where $k=1,\dots,\mu$, i.e. the projections of $\rho(0)$ onto a set of operators $\{Q_i \circ A_1^T, \dots, Q_i \circ A_{\mu}^T \}$. Let us denote the subspace spanned by these operators by $\mathcal{S}(Q_i; A_1, \dots, A_{\mu})$, i.e.
\begin{equation}\label{25}
\mathcal{S}(Q_i; A_1, \dots, A_{\mu}) \equiv Span \{Q_i \circ A_1^T, \dots, Q_i \circ A_{\mu}^T\}.
\end{equation}

By performing at time instants $\{t_1, \dots, t_p\}$ the measurement of each observable from a set $\{Q_1, \dots, Q_r\}$ we obtain a matrix of data $[Tr\left\{ (Q_i \circ A_k^T) \rho(0) \right\}]$ where $i=1,\dots,r$ and $k=1,\dots,\mu$. The condition for reconstructability of the initial state based on this set of data can be formulated by means of the subspace introduced in \eqref{25}, analogously as in case of known results of the stroboscopic tomography \cite{jam83,jam12}.

\begin{thm}\label{thm3}
The quantum state $\rho(0)$ is reconstructible from the projections $[Tr\left\{ (Q_i \circ A_k^T) \rho(0) \right\}]$ where $i=1,\dots,r$ and $k=1,\dots,\mu$ if and only if
\begin{equation}\label{26}
\bigoplus_{i=1}^r \mathcal{S}(Q_i; A_1, \dots, A_{\mu}) = \mathcal{B(H)},
\end{equation}
\end{thm}
where $\bigoplus$ denotes the Minkowski sum of subspaces.

The equation \eqref{26} can be put into words and stated that the set of operators $\{Q_i \circ A_k^T\}$ where $ i=1,\dots,r$ and $k=1,\dots,\mu$, i.e. these are the operators onto which $\rho(0)$ is projected, have to span the space to which $\rho(0)$ belongs. In other words, the set $\{Q_i \circ A_k^T\}$ where $ i=1,\dots,r$ and $k=1,\dots,\mu$ has to be a spanning set. The condition expressed in theorem \ref{thm3} represents another way to state that a set is complete. One can compare theorem \ref{thm3} with four equivalent definitions of spanning set given in \cite{ariano2000} -- one shall see that the meaning is the same, only the terminology is different.

General theorems and observations concerning quantum tomography for phase-damping channels can be employed to solve a huge number of research problems. Nevertheless, in this analysis we shall demonstrate how this approach to quantum tomography works on two specific examples.

\section{An Example for $dim\mathcal{H} =2$ -- Dephasing}\label{sec4}

In order to demonstrate how our approach to quantum tomography works, let us consider an open quantum system associated with the Hilbert space $\mathcal{H}$ such that $dim \mathcal{H} =2$. The evolution of this quantum system is given by a Kossakowski-Lindblad master equation of the form:
\begin{equation}\label{27}
\mathbb{L}[\rho] = \frac{\gamma}{2} (\sigma_3 \rho \sigma_3 - \rho),
\end{equation}
where $\sigma_3 = \left( \begin{matrix} 1 & 0 \\ 0 &-1 \end{matrix} \right)$ and $\gamma>0$ is a parameter.

This evolution model is often referred to as dephasing. To observe how a quantum system subject to such dynamic changes in time let us introduce the following notation for the initial density matrix
\begin{equation}\label{28}
\rho(0) = \left [ \begin{matrix} \rho_{11} & \rho_{12} \\ \rho_{21} & \rho_{22} \end{matrix} \right].
\end{equation}

Naturally, $\rho(t)$ can be computed by solving the differential equation \eqref{27}:
\begin{equation}\label{29}
\rho(t) = exp(\mathbb{L}t) \rho(0).
\end{equation}

One can obtain a specific form of the solution 
\begin{equation}\label{30}
\rho(t) = \left [ \begin{matrix} \rho_{11} & e^{-\gamma t} \rho_{12} \\ e^{- \gamma t} \rho_{21} & \rho_{22} \end{matrix} \right]
\end{equation}
and rewrite it by means of the Hadamard product
\begin{equation}\label{31}
\rho(t) = \left [ \begin{matrix} 1 & e^{-\gamma t} \\ e^{- \gamma t}  & 1 \end{matrix} \right] \circ \left [ \begin{matrix} \rho_{11} &  \rho_{12} \\ \rho_{21} & \rho_{22} \end{matrix} \right],
\end{equation}
which indicates that this kind of evolution belongs to the class of phase-damping channels (or equivalently pure decoherence) and, therefore, the approach introduced in \ref{sec3} is applicable.

Let us denote analogously as earlier
\begin{equation}\label{32}
D(t) \equiv \left [ \begin{matrix} 1 & e^{-\gamma t} \\ e^{- \gamma t}  & 1 \end{matrix} \right],
\end{equation}
then one can find the decomposition of $D(t)$ as introduced in \eqref{decom} in the form
\begin{equation}\label{33}
D(t) = A_1 + e^{- \gamma t} A_2,
\end{equation}
where $A_1 = \left [ \begin{matrix} 1 & 0 \\ 0  & 1 \end{matrix} \right]$ and $A_2 = \left [ \begin{matrix} 0 & 1 \\ 1  & 0 \end{matrix} \right]$.

Let us assume that from an experiment we can obtain mean values of two observables $Q_1$ and $Q_2$ of the form
\begin{equation}\label{34}
Q_1 = \sigma_1 = \left [ \begin{matrix} 0 & 1 \\ 1  & 0 \end{matrix} \right] \textit{ } Q_2 = \sigma_2 + \sigma_3 = \left [ \begin{matrix} 1 & -i \\ i  & -1 \end{matrix} \right].
\end{equation}

One can notice that $Tr\{ (Q_1 \circ A_1^T) \rho(0)\} = 0$ and $Tr\{ (Q_1 \circ A_2^T) \rho(0)\} = Tr\{\sigma_1 \rho(0)\}$, which means that if one measures $Q_1$ at one time instant, one can get the projection of $\rho(0)$ onto $\sigma_1$ because
\begin{equation}\label{35}
m_1 (t) = Tr\{ (Q_1 \circ A_1^T) \rho(0)\} + e^{- \gamma t}Tr\{ (Q_1 \circ A_2^T) \rho(0)\}  = Tr\{Q_1 (D(t) \circ \rho(0))\} = e^{- \gamma t} Tr\{ \sigma_1 \rho(0)\}. 
\end{equation}

Without any loss of generality we can assume that $t=0$, which simplifies the considerations in equation \eqref{35} to a form
\begin{equation}\label{36}
m_1(0) = Tr\{ \sigma_1 \rho(0)\}.
\end{equation}

In case of the observable $Q_2$, one can notice that $Tr\{ (Q_2 \circ A_1^T) \rho(0)\} = Tr\{\sigma_3 \rho(0)\}$ and $Tr\{ (Q_2 \circ A_2^T) \rho(0)\}=  Tr\{\sigma_2 \rho(0)\}$, which means that if one measures the observable $Q_2$ at two time instants $t_1,t_2 \geq 0$ obtaining values $m_2(t_1), m_2(t_2)$, one can get a matrix equation in the form
\begin{equation}\label{37}
\left [ \begin{matrix} m_2(t_1) \\ m_2(t_2) \end{matrix} \right] = \left [ \begin{matrix} 1 & e^{- \gamma t_1} \\ 1  & e^{- \gamma t_2} \end{matrix} \right] \left [ \begin{matrix} Tr\{\sigma_3 \rho(0)\} \\ Tr\{\sigma_2 \rho(0)\} \end{matrix} \right].
\end{equation}

Obviously, if the time instants satisfy the condition $t_1 \neq t_2$, one can calculate from the equation \eqref{37} the projections $Tr\{\sigma_3 \rho(0)\}$ and $ Tr\{\sigma_2 \rho(0)\}$. For simplicity we can assume that the $t_1=0$ and $t_2 = t>0$. Then one can obtain:
\begin{equation}\label{38}
Tr\{\sigma_3 \rho(0)\} = \frac{m_2(0) e^{-\gamma t} - m_2(t)}{e^{-  \gamma t} -1} \text{ , } Tr\{\sigma_2 \rho(0)\} = \frac{m_2(t) -m_2(0)}{e^{- \gamma t} -1}.
\end{equation}

Bearing in mind that for $dim\mathcal{H}=2$ any density matrix can be decomposed as
\begin{equation}\label{39}
\rho = \frac{1}{2} \left( \mathbb{I}_2 + \sum_{i=1}^3 Tr\{\sigma_i \rho\} \sigma_i \right),
\end{equation}
one obtains an explicit formula for the initial density matrix of a quantum system with evolution given by \eqref{27}:
\begin{equation}\label{40}
\rho(0) = \frac{1}{2} \left( \mathbb{I}_2 + m_1(0) \sigma_1 + \frac{m_2(t) -m_2(0)}{e^{- \gamma t} -1} \sigma_2 + \frac{m_2(0) e^{-\gamma t} - m_2(t)}{e^{-  \gamma t} -1} \sigma_3 \right ),
\end{equation}

where $\mathbb{I}_2 = \left [ \begin{matrix} 1 & 0 \\ 0 & 1 \end{matrix} \right]$.

Apparently the result presented in \eqref{39} is the final result of the quantum tomography model, because the unknown initial density matrix have been expressed by means of accessible data -- the values $m_1(0), m_2(0),m_2(t)$ are assumed to be available from an experiment, $t$ can be any number greater than zero and $\gamma$ is a parameter which depends on the character of evolution. If one can reconstruct the initial density matrix and one has the knowledge about the evolution, then one is able to determine the complete trajectory of the quantum state.

One can notice that the result \eqref{39} is equivalent to the result concerning dephasing published in \cite{czerwin15}, where the author followed the stroboscopic approach as introduced in \cite{jam83}. The fact that both results are practically identical confirms that the new approach outlined in the current article is correct. However, if one compares the current dynamic tomography model based on dynamical maps given by Hadamard product with the stroboscopic approach as in \cite{jam83}, one can notice that in the current analysis we do not need to refer to the notion of the minimal polynomial of the generator of evolution, which is unavoidable in case of the stroboscopic tomography. For higher dimensional cases it is usually difficult to study the algebraic properties of the generator of evolution. Therefore, the current approach may have an advantage over the stroboscopic tomography for cases when the dynamical map can be given by the Hadamard product.

\section{An Example for $dim\mathcal{H} =N$ -- Gaussian Semigroups}\label{sec5}

The example solved in the previous section is rather simple and, therefore, in this part we shall analyze a more complex example of an open quantum system connected with $N-$dimensional Hilbert space. In our example the dynamics of the system is given by a master equation of the following form:

\begin{equation}\label{gaussian}
\frac{d \rho(t)}{d t} = - \frac{1}{2} [H, [H, \rho (t) ]],
\end{equation}

where $H \in \mathcal{B_* (H)}$.

The initial density matrix can be decomposd in the basis which consists of the eigenvectors of $H$, i.e.

\begin{equation}
\rho (0) = \sum_{n,m} \rho_{nm} | \phi_n \rangle \langle \phi_m|,
\end{equation}

where vectors $|\phi_n\rangle$ satisfy $H |\phi_n \rangle = E_n |\phi_n \rangle$.

One can easily find the solution to the equation \eqref{gaussian}:

\begin{equation}
\rho(t) = \sum_{n,m} \rho_{nm}  e^{-\frac{1}{2} t (E_n - E_m)^2} | \phi_n \rangle \langle \phi_m|
\end{equation}

From this solution one can see the structure of the matrix $D(t)$ such that $\rho (t) = D(t) \circ \rho(0)$:

\begin{equation}
d_{nm} (t) = e^{-\frac{1}{2} t (E_n - E_m)^2}.
\end{equation}

In order to analyze the properties of the matrix $D(t)$ one needs to make some assumptions about the eigenvalues of $H$. Naturally, within the spectrum of $H$ there might be some degeneracies and, therefore, we shall assume that there are $\xi$ distinct eigenvalues and we shall denote them as $\lambda_1, \dots, \lambda_{\xi}$. Moreover, we shall assign to each distinct eigenvalue its multiplicity denoted by $n_1, \dots, n_{\xi}$. Finally, we shall assume that the eigenvalues constitute an arithmetic progression, i.e. $\lambda_k - \lambda_{k-1} = c$ for $k=2, \dots, \xi.$

Under such conditions we can formulate a theorem concerning the minimal number of distinct observables required for quantum tomography of systems with evolution given by \eqref{gaussian}.

\begin{thm}
The minimal number of distinct observables required for quantum tomography of systems with evolution given by \eqref{gaussian} can be expressed by the formula
\begin{equation}
\eta = max \{ \beta_0, \beta_1, \dots, \beta_r \},
\end{equation}
where $r = \frac{\xi-1}{2}$ if $\xi$ is odd or $r=\frac{\xi-2}{2}$ if $\xi$ is even. The values of $\beta_0,\beta_1, \dots, \beta_r$ are computed from equations:
\begin{equation}
\beta_0 := n_1^2 + n_2^2 + \dots + n_{\xi}^2,
\end{equation}

\begin{equation}
\beta_k := 2 \sum_{i=1}^{\xi-k} n_i n_{i+k}.
\end{equation}
\end{thm}

\begin{proof}

The matrix $D(t)$ has the following block form:

\begin{equation}
D(t) =\left[\begin{array}{c|c|c|c|c}
\tilde{D}_{11} (t) & \tilde{D}_{12} (t) & \tilde{D}_{13} (t) & \dots & \tilde{D}_{1\xi} (t)   \\
\hline
\tilde{D}_{21} (t) & \tilde{D}_{22} (t) & \tilde{D}_{23} (t) & \dots & \tilde{D}_{2\xi} (t) \\
\hline
\tilde{D}_{31} (t) & \tilde{D}_{32} (t) & \tilde{D}_{33} (t) & \dots & \tilde{D}_{3\xi} (t) \\
\hline
\vdots & \vdots & \vdots & \ddots & \vdots \\
\hline
\tilde{D}_{\xi 1} (t) & \tilde{D}_{\xi 2} (t) & \tilde{D}_{\xi 3} (t) & \dots & \tilde{D}_{\xi \xi} (t) \\
\end{array}
\right],
\end{equation}

where each block $\tilde{D}_{ij}(t)$ is a $n_i \times n_j$-dimensional matrix with all the elements equal to one another and given by $\tilde{d}_{ij}^{qp} (t) = e^{-\frac{1}{2} t ((i-j)c)^2},$ where $q=1,\dots, n_i$ and $p=1, \dots, n_j$. One can obtain the decomposition of $D(t)$ in the form of \eqref{decom}:

\begin{equation}
D(t) = \sum_{k=0}^{\xi -1} e^{-\frac{1}{2} t (kc)^2} A_k.
\end{equation}

By $A_0, A_1, \dots, A_{\xi-1}$ we denote the following matrices
\begin{equation}
A_0 =\left[\begin{array}{c|c|c|c|c}
\tilde{J}_{11} &  &  & \dots & \\
\hline
 & \tilde{J}_{22} &  & \dots & \\
\hline
 &   & \tilde{J}_{33} & \dots &  \\
\hline
\vdots & \vdots & \vdots & \ddots & \vdots \\
\hline
 &  &  & \dots & \tilde{J}_{\xi \xi} \\
\end{array}
\right],
\end{equation}

\begin{equation}
A_1 =\left[\begin{array}{c|c|c|c|c}
& \tilde{J}_{12}  &  & \dots & \\
\hline
 \tilde{J}_{21} & &\tilde{J}_{23}  & \dots & \\
\hline
 & \tilde{J}_{32}  &  & \dots &  \\
\hline
\vdots & \vdots & \vdots & \ddots & \vdots \\
\hline
 &  &  & \dots &  \\
\end{array}
\right],
\end{equation}
\begin{equation}
A_{\xi-1} =\left[\begin{array}{c|c|c|c|c}
&   &  & \dots & \tilde{J}_{1 \xi}\\
\hline
  & &  & \dots & \\
\hline
 &   &  & \dots &  \\
\hline
\vdots & \vdots & \vdots & \ddots & \vdots \\
\hline
\tilde{J}_{\xi 1} &  &  & \dots &  \\
\end{array}
\right],
\end{equation}

where the symbol $\tilde{J}_{i j}$ denotes a $n_i \times n_j$-dimensional matrix with all the elements equal to one. Empty blocks in those matrices mean that there are only zeros.

Now, as we have the desired decomposition of $D(t)$, we can post a question: \textit{What is the minimal number of observables required for quantum tomography of system with evolution given by} \eqref{gaussian}?

If one measures a certain observable $Q_i$ at some time instant $t_j$, one can represent the measurement result as

\begin{equation}\label{measurement}
m_i (t_j) = \sum_{k=0}^{\xi  -1} e^{-\frac{1}{2} t (kc)^2} Tr\left\{ (Q_i \circ A_k) \rho(0)\right\}.
\end{equation}

As it was explained in the previous sections, from the equations \eqref{measurement} we can calculate the projections $Tr\left\{ (Q_i \circ A_k)\rho(0)\right\}$ where $k =1,\dots, \xi-1$. Let us find the minimal number of observables $Q_1,Q_2, \dots,Q_{\eta}$ such that one can reconstruct the initial density matrix $\rho(0)$ on the basis of the pojections $Tr\left\{ (Q_i \circ A_k)\rho(0)\right\}$ where $k =1,\dots, \xi-1$ and $i=1,\dots, \eta$.

Let us first consider products $Tr\left\{ (Q_i \circ A_0)\rho(0)\right\}$ where $i=1,\dots, \eta$. Such projections should provide knowledge about the blocks of $\rho(0)$ which correspond with the non-zero blocks of $A_0$. Therefore, we need at least $n_1^2 +n_2^2+ \dots + n_{\xi}^2 \equiv \beta_0 $ observables. 

If one considers products $Tr\left\{ (Q_i \circ A_1)\rho(0)\right\}$ where $i=1,\dots, \eta$. Then such projections should provide knowledge about the blocks of $\rho(0)$ which correspond with the non-zero blocks of $A_1$. Therefore, we need at least $2n_1 n_2 + 2n_2 n_3 + \dots 2 n_{\xi -1} n_{\xi} \equiv \beta_1 $ observables.

If one considers products $Tr\left\{ (Q_i \circ A_2)\rho(0)\right\}$ where $i=1,\dots, \eta$. Then such projections should provide knowledge about the blocks of $\rho(0)$ which correspond with the non-zero blocks of $A_2$. Therefore, we need at least $2n_1 n_3 + 2n_2 n_4 + \dots 2 n_{\xi -2} n_{\xi} \equiv \beta_2 $ observables. 

This reasoning is worth continuing till calculating $\beta_r$ where $r = \frac{\xi-1}{2}$ if $\xi$ is odd or $r=\frac{\xi -2}{2}$ if $\xi$ is even. Finally, out of the numbers $\beta_0, \beta_1, \dots, \beta_r$ one needs to choose the greates value, which completes the proof.

\end{proof}

\section{Conclusion}

This article introduces a dynamic approach to quantum tomography of systems with evolution given by phase-damping channels. We have presented a general method for state reconstruction of systems subject to phase-damping along with an example that shows how to create a complete quantum tomography model based on this method. This kind of evolution, often also referred to as "pure decoherence", appears in natural considerations in the theory of open quantum system. Therefore, we believe that the content of the current article is relevant to modern physical theories. We have proved that knowledge about the evolution combined with selected algebraic properties can bring a new insight into quantum tomography. If one is thinking of implementing quantum tomography concepts in experiments, he or she will find it advantageous to put into practice dynamic tomography model, because it allows to reconstruct the quantum state in an economical way, i.e. starting with an informationally incomplete set of observables one can perform the measurement of each observable at some distinct time instants, which may lead to obtaining sufficient data for state reconstruction. In case of phase-damping channels there is a need for more research concerning optimal criteria for quantum tomography, i.e. minimal number of distinct observables expressed by properties of $D(t)$ from \eqref{3} and an optimal way to decompose $D(t)$ as in \eqref{decom}. Such question shall be considered in forthcoming articles.

\section*{Acknowledgement}
This research has been supported by the grant No. 2015/17/B/ST2/02026 of National Science Center of Poland.

\end{document}